\documentclass{iopbk2e}
\usepackage{iopams,graphicx}

\title{3rd Generation Photovoltaics for High Efficiency
through Full Spectrum Utilization}

\author{\emph{Editors:}\\A Luque and A Marti}

\newcommand{\ingaasp}{$\rm In_{1-x}\-Ga_{x}\-As_{y}\-P_{1-y}$}
\newcommand{\ingaas}{$\rm In_{1-x}\-Ga_{x}\-As$}
\newcommand{\bulkingaas}{$\rm In_{0.53}\-Ga_{0.47}\-As$}
\newcommand{\ingaasingaas}{\ingaas/$\rm In_{1-z}\-Ga_{z}\-As$}

\newcommand{\jsc}{$J_{sc}$\ }
\newcommand{\voc}{$V_{oc}$\ }

\begin{document}

%\maketitle

\chapter[QWs in PV]{Quantum Wells in Photovoltaic Cells}
{C Rohr, P Abbott, I M Ballard, D B Bushnell,
J P Connolly, N J Ekins-Daukes, K W J Barnham\\
Experimental Solid State Physics, Imperial College London, U.K.}

\section{Introduction}\label{sec:intro}
The fundamental efficiency limit of a single bandgap solar cell is
about 31\% at one sun with a bandgap of about E$_g = 1.35$ eV
\cite{henry:80}, determined by the trade-off of maximising current
with a smaller bandgap and voltage with a larger bandgap. Multiple
bandgaps can be introduced to absorb the broad solar spectrum more
efficiently. This can be realised in multi-junction cells, for
example, where two or more cells are stacked on top of each other
either mechanically or monolithically connected by a tunnel
junction. An alternative -- or complementary (see
\sref{sec:tandem}) -- approach is the quantum well cell (QWC).

\section{\index{Quantum Well Cells}}\label{sec:qwc}

\begin{figure}
\center \includegraphics*[width=100mm]{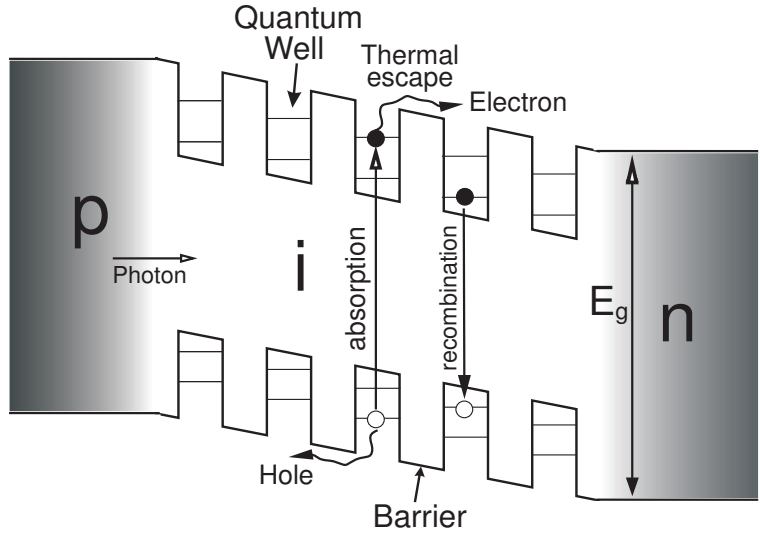}
\caption{Schematic bandgap diagram of a Quantum Well Cell. The
absorption threshold is determined by the lowest energy levels in
the quantum wells. Carriers escape thermally assisted and by
tunnelling.} \label{MQW-schematic}
\end{figure}

\index{Quantum wells} (QWs) are thin layers of lower bandgap
material in a host material with a higher bandgap. Early device
designs placed the QWs in the doped regions of a p-n device
\cite{chaffin:84}, but superior carrier collection is achieved
when an electric field is present across the QWs.  More recent QWC
designs have employed a p-i-n structure \cite{barnham:90} with the
QWs located in the intrinsic region; a schematic bandgap diagram
is shown in \fref{MQW-schematic}. The carriers escape from the QWs
thermally and by tunnelling
\cite{nelson:93,barnes:97,zachariou:98}.

The photocurrent is enhanced in a QWC compared to a cell made
without QWs also known as barrier control, and experimentally it
is observed that the voltage is enhanced compared with a bulk cell
made of the QW material \cite{barnham:96}. Hence QWCs can enhance
the efficiency if the photocurrent enhancement is greater than the
loss in voltage \cite{barnham:97}. The potential for an efficiency
enhancement is also discussed in references
\cite{anderson:02,honsberg:02}. The number of QWs is limited by
the maximum thickness of the i-region maintaining an electric
field across it. QWCs have been investigated quite extensively,
both on GaAs as well as on InP substrates, and have been discussed
in some detail also in references \cite{nelson:95tf,nelson:01}.

\begin{figure}
\center \includegraphics*[width=110mm]{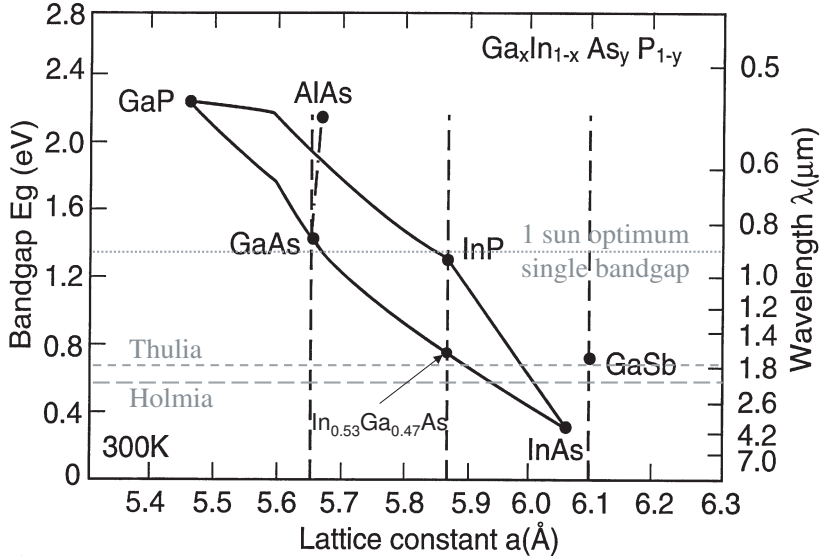}
\caption{Lattice constant versus bandgap for \ingaasp\ and AlGaAs
compounds. Also indicated is the optimum bandgap for a
single-bandgap PV cell under 1 sun, and the emission peaks of
selective emitters Thulia and Holmia.} \label{Eg-a}
\end{figure}

Historically, the first p-i-n QWCs were in the material system
AlGaAs/GaAs (barrier/well) on GaAs
\cite{barnham:91,fox:93,paxman:93,aperathitis:98,connolly:98}.
AlGaAs is closely lattice-matched to GaAs and the bandgap can be
easily varied by changing the Al fraction (see \fref{Eg-a}) up to
about 0.7 where the bandgap becomes indirect. However the material
quality particularly that of AlGaAs is relatively poor because of
contamination during the epitaxial growth, leading to a high
number of recombination centres and hence a high dark current.

An alternative material to AlGaAs is InGaP which has better
material quality, and an InGaP/GaAs QWC has been demonstrated
\cite{zachariou:v98}. However, the ideal single bandgap for a 1
sun solar spectrum is E$_g = 1.35$ eV, as indicated in
\fref{Eg-a}, while GaAs has a bandgap of E$_g = 1.42$ eV and that
of AlGaAs is still higher. The second material should therefore
have a smaller bandgap than GaAs to absorb the longer wavelength
light, keeping in mind that the quantum confinement raises the
effective bandgap of the QWs.

GaAs/InGaAs QWCs fulfil this criterion and they have been studied
quite extensively
\cite{barnes:94,ragay:94,barnes:96,griffin:96,freundlich:98,ekins:rev01}.
However, because InGaAs has a larger lattice constant than GaAs
(see \fref{Eg-a}) it is strained. If the strain exceeds a critical
value relaxation occurs at the top and bottom of the MQW stack,
and the dislocations result in an increase in recombination and
hence increased dark current \cite{griffin:96}. This limitation
means that strained GaAs/InGaAs QWCs cannot improve the efficiency
compared to GaAs control cells \cite{ekins:rev01}.

Strain compensation techniques can be used to overcome this
problem (see \sref{sec:sb}), and QWCs in the material system
GaAsP/InGaAs on GaAs have been investigated
\cite{ekins:apl99,ekins:rev01,ekins:00,bushnell:00,ekins:01,bushnell:01}.
These devices are also very suitable as bottom cells in a tandem
configuration (see \sref{sec:sb}).

QWCs based on InP are of interest for solar as well as for
thermo-photovoltaic (TPV) applications. First, material
combinations such as InP/InGaAs were investigated
\cite{freundlich:94,zachariou:96,griffin:97,zachariou:98}, which
was then extended to quaternary material (lattice-matched to InP)
InP/\ingaasp\ ($x=0.47y$)
\cite{griffin:98,rohr:nrel:98,raisky:98}.

As in the GaAsP/InGaAs system on GaAs, strain compensation
techniques have been employed in \ingaasingaas\ QWCs on InP
\cite{rohr:99,rohr:00,rohr:02,abbott:02,rohr:tpv02,abbott:tpv02}.

QWCs have practical advantages due to both quantised energy levels
and the greater flexibility of choice of materials. In particular,
this allows engineering of the bandgap to better match the
incident spectrum. The absorption threshold can be varied by
changing the width of the QW and/or by changing its material
composition.

This flexibility can be further increased by employing strain
compensation techniques which are explained in more detail in
Section~\sref{sec:sb}. In this way, longer wavelengths for
absorption can be achieved than what is possible with
lattice-matched bulk material, allowing optimisation of the
bandgap.

The application of QWCs (based on InP) for thermophotovoltaics is
discussed in \sref{sec:tpv}. For TPV applications the same concept
of strain compensation can be applied to extend the absorption to
longer wavelengths. This is important for relatively low
temperature sources combined with appropriate selective emitters
for example based on Holmia or Thulia.

Several studies indicate that QWCs have a better temperature
dependence of efficiency than bulk cells
\cite{aperathitis:98,ballard:phd,rohr:phd,ballard:01}.

\section{Strain compensation}\label{sec:sb}

\begin{figure}
\center \includegraphics*[width=85mm]{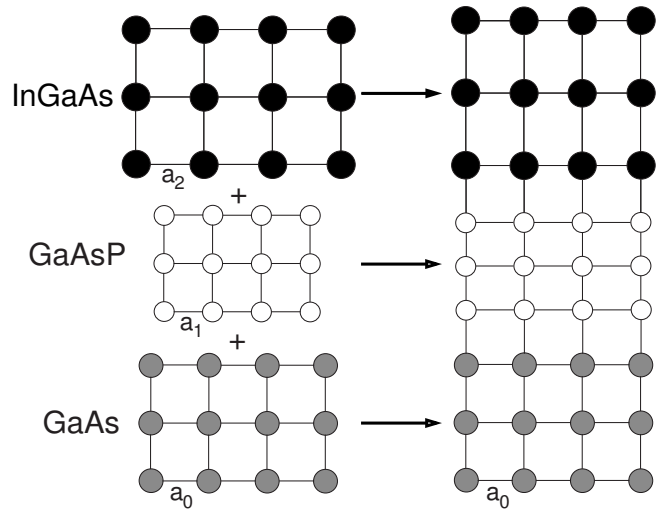}
\caption{Schematic diagram of strain compensation: the natural
lattice constant of GaAsP (a$_1$) is smaller than that of the GaAs
substrate (a$_0$), and GaAsP barriers are therefore tensile
strained, while the natural lattice constant of InGaAs (a$_2$) is
larger, and hence InGaAs QWs are compressively strained.}
\label{strainbalance}
\end{figure}

In order to avoid strain relaxation, strain compensation
techniques, first proposed by Matthews and Blakeslee
\cite{matthews:76}, can be used to minimise the stress at the
interface between the substrate and a repeat unit of two layers
with different natural lattice constants. Layers with larger and
smaller natural lattice constant compared to the substrate result
in compressive and tensile strain respectively as shown in
\fref{strainbalance}. When these two layers are strained against
each other, the strain is compensated and the net force exerted on
the adjacent layers is reduced. Therefore the build up of strain
in a stack can be reduced, and hence its critical thickness is
increased so that more such repeat units can be grown on top of
each other without relaxation. If the strain compensation
conditions are optimised to give zero stress at the interfaces
between the repeat units, an unlimited number of periods can be
grown in principle. In addition each individual layer has to
remain below its critical thickness which means that this concept
can only be used for thin layers such as quantum wells.

\begin{figure}
\center \includegraphics*[width=60mm]{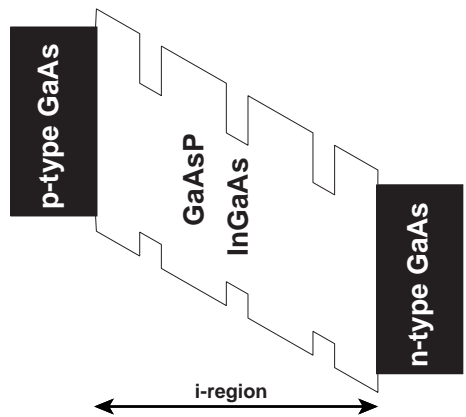}
\caption{Bandgap diagram of a strain compensated QWC with InGaAs
QWs and GaAsP barriers.} \label{strainbalance-bandgap}
\end{figure}

This technique is very suitable for multi quantum well structures;
the barriers and wells are made of different materials with larger
and smaller bandgaps but they can also have smaller and larger
lattice constants (see \fref{Eg-a}), i.e.\ with tensile and
compressive strain respectively (see \fref{strainbalance} and
\fref{strainbalance-bandgap}). That means strained materials can
be used for the quantum wells in order to reach lower bandgaps
without compromising the quality of the device as dislocations are
avoided. This technique, which extends the material range allowing
further bandgap engineering, was first applied to photovoltaics in
GaAsP/InGaAs QWCs \cite{ekins:apl99}.

For highly strained layers the difference in elastic constants
becomes significant and it needs to be taken into account when
considering the conditions for zero stress \cite{ekins:02}. The
strain for each layer $i$\/ is
%defined as
\begin{equation}\label{strain}
    \epsilon_i = \frac{a_0 - a_i}{a_i}
\end{equation}
where $a_0$ is the lattice constant of the substrate and $a_i$ the
natural lattice constant of layer $i$. The zero-stress
strain-balance conditions are as follows:
\begin{eqnarray}
    \epsilon_1 t_1 A_1 a_2 + \epsilon_2 t_2 A_2 a_1 = 0 \\
    a_0 = \frac{t_1 A_1 a_1 a_2^2 + t_2 A_2 a_2 a_1^2}{t_1 A_1 a_2^2 + t_2 A_2 a_1^2}
\end{eqnarray}
where $A=C_{11} + C_{12} - \frac{2 C_{12}^2}{C_{11}}$  with
elastic stiffness constants $C_{11}$ and $C_{12}$, different for
each layer, depending on the material.

The strain energy must be kept below a critical value, however, to
avoid the onset of three-dimensional growth \cite{nasi:02}.
Lateral layer thickness modulations, particularly in the tensile
strained material (i.e.\ barriers), are origins of dislocations
and result in isolated highly defected regions if the elastic
strain energy density reaches a critical value. In practice the
strain balancing puts stringent requirements on growth.

\section{QWs in Tandem Cells}\label{sec:tandem}

\begin{figure}
\center \includegraphics*[width=110mm]{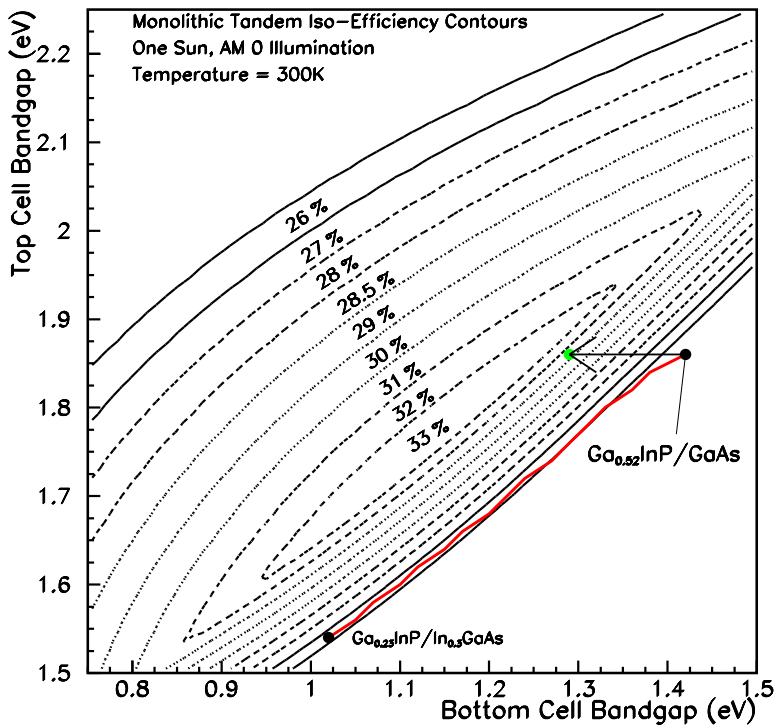}
\caption{Contour plot of tandem cell efficiency as a function of
top and bottom cell bandgaps.} \label{fan}
\end{figure}

In tandem cells two photovoltaic cells with two different bandgaps
are stacked on top of each other. The bandgaps of the two cells
can be optimised for the solar spectrum \cite{fan:82}, and a
contour plot of efficiency as a function of top and bottom cell
bandgaps is shown in \fref{fan} for an AM0 spectrum. Tandem cells
can be grown monolithically on a single substrate, connected with
a tunnel junction. The top and bottom cell are connected in
series, which means that the lower photocurrent of the two cells
determines the current of the tandem device. Monolithic growth
requires that the materials are lattice matched in order to avoid
relaxation. A tandem with a GaInP top cell lattice matched to a
GaAs bottom cell does not have the optimum bandgap combination as
one can see in \fref{fan}, and there is no good quality material
available with a smaller bandgap than GaAs having the same lattice
constant.

The standard approach to matching the currents in the GaInP/GaAs
tandem is by thinning the top cell so that enough light is
transmitted to the bottom cell to generate more current there
\cite{kurtz:90b}. This is not an optimum configuration, however,
as the quantum efficiency of the top cell is reduced and in
addition there are losses in the bottom cell due to more
high-energy carriers relaxing to the bandedge.

It is possible to grow a virtual substrate, where the lattice
constant is relaxed and the misfit dislocations are largely
confined to electrically inactive regions of the device. In this
way the lattice constant can be changed before growing the tandem.
But the bottom and top cell of the tandem should still be lattice
matched with respect to each other and hence the combination of
bandgaps is restricted indicated with a line in \fref{fan}; the
optimum where the currents in both cells are matched cannot be
reached in general. Several groups have grown samples that fall on
that line in \fref{fan} \cite{hoffman:98,dimroth:01b}.

QWs extend the absorption and therefore increase the photocurrent
compared with a barrier control; hence a GaAs/InGaAs QWC generates
more current than a GaAs cell and can be better matched to a GaInP
top cell in a tandem cell under the solar spectrum
\cite{connolly:98,freundlich:98,freundlich:98patent,freundlich:02patent}.
However, as mentioned in \sref{sec:qwc}, the voltage deteriorates
due to the formation of misfit dislocations with increasing strain
when incorporating strained InGaAs QWs \cite{griffin:96}.

Strain compensated QWCs offer an attractive solution in that the
link between lattice matching and bandgap is decoupled
\cite{ekins:rev01}. The bandgap of the bottom cell, for example,
can be reduced with a strain compensated QWC, which means that one
can move along a horizontal line in \fref{fan} improving the
efficiency of a tandem cell quite rapidly \cite{ekins:apl99}. QWCs
for both the top and the bottom cell give an extra degree of
freedom and optimisation of the bandgaps to obtain maximum
efficiency becomes possible.

\section{QWCs with light trapping}\label{sec:dbr}

Not all the light is absorbed in the quantum wells because they
are optically thin and their number limited. Hence light trapping
techniques to increase the number of light passes through the MQW
is desirable, boosting the QW photo-response significantly. The
simplest form is a mirror on the back surface resulting in two
light passes. Texturing the front or the back surface can further
increase the path length of the light, in particular if the light
is at a sufficiently large angle with respect to the normal so to
obtain internal reflections. Another option is to incorporate
gratings into the structure to diffract the light to large angles
\cite{bushnell:02}.

Light trapping is only desirable for the wavelength range where
the optically thin QWs absorb. Other parts of the cell are
optically thick and hence light trapping for photons with energies
greater than the bandgap of the bulk material has a minimal effect
if any. The energy of photons that are trapped but not absorbed
(e.g.\ below bandgap photons) and additional high-energy carriers
relaxing to the bandedge contribute to cell heating and are
therefore undesirable.

A solution to this problem is to use a wavelength specific mirror
such as a distributed Bragg reflector (DBR) \cite{bushnell:01}. A
DBR consists of alternating layers of high and low refractive
index material, each one-quarter of a wavelength thick.
Constructive interference occurs for the reflected light of the
design wavelength and adding periods to the DBR causes a higher
reflection due to the presence of more in-phase reflections from
the added interfaces. DBRs maintain a high reflectance within a
region around the design wavelength known as the stop-band. As the
refractive index contrast between the two materials in the DBR
increases, the peak reflection rises and the stop-band widens.

\begin{figure}
\center \includegraphics*[width=110mm]{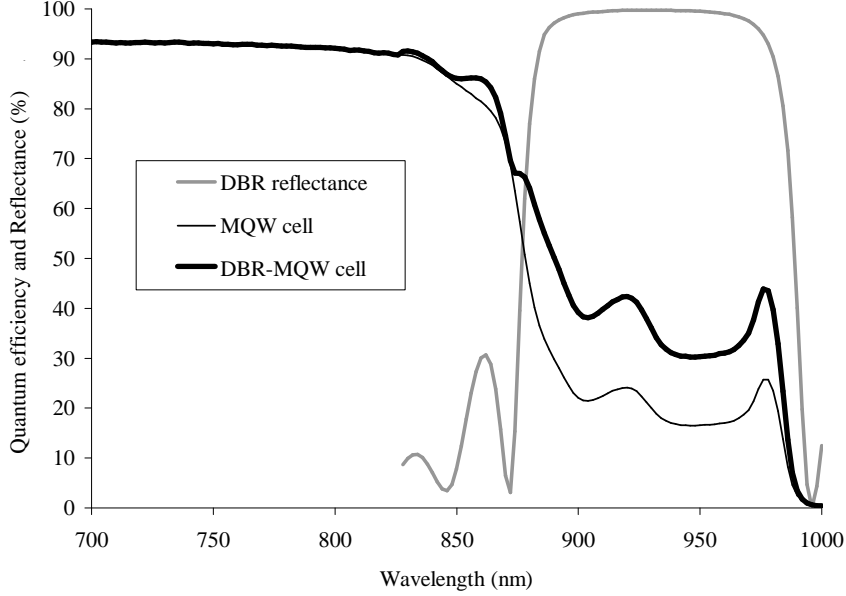}
\caption{Quantum efficiency of a 20 QW device with and without
DBR, and reflectance of the DBR.} \label{dbr}
\end{figure}

This technique is particularly attractive for multi-junction cells
where the second cell is a QWC with a DBR on an active Ge
substrate. In \fref{dbr} the quantum efficiency of a typical QWC
with 20 QW is shown with and without a 20.5 period DBR, as well as
the reflectance of the DBR; the sample description of the QWC with
DBR is given in \tref{tab:qwc-dbr}. The quantum efficieny (QE) of
the QWC with DBR in \fref{dbr} is calculated from the measured QE
and the calculated DBR reflectance. The DBR reflects back only the
relevant wavelength range that can be absorbed in the MQW,
significantly boosting its spectral response.

\begin{table}
\caption{Sample description of QWC with DBR.} \label{tab:qwc-dbr}
\begin{center}
\begin{tabular}{@{}lccccc@{}}
\br Layer & Repeats & Material & Thickness (\AA) & Doping & Conc. (cm$^{-3}$)\\
\mr Cap & \01 & GaAs & \02200 & p & $2 \times 10^{19}$ \\
Window & \01 & Al$_{0.8}$GaAs & \0\0450 & p & $1 \times 10^{18}$ \\
Emitter & \01 & GaAs & \05000 & p & $2 \times 10^{18}$ \\
i-region & \01 & GaAs & \0\0100 & &  \\
$\frac{1}{2}$ barrier & 20 & GaAsP$_{0.06}$ & \0\0210 & & \\
QW & 20 & In$_{0.17}$GaAs& \0\0\070 & & \\
$\frac{1}{2}$ barrier & 20 & GaAsP$_{0.06}$ & \0\0210 & & \\
i-region & \01 & GaAs & \0\0100 & & \\
Base & \01 & GaAs & 24000 & n & \0$1.5\times 10^{17}$ \\
DBR & 20  & Al$_{0.5}$GaAs & \0\0144 & n & $1 \times 10^{18}$ \\
DBR & 20 & AlAs & \0\0538 & n & $1 \times 10^{18}$\\
DBR & 20 & Al$_{0.5}$GaAs & \0\0144 & n & $1 \times 10^{18}$\\
DBR & 20 & Al$_{0.13}$GaAs & \0\0520 & n & $1 \times 10^{18}$\\
DBR & \01 & Al$_{0.5}$GaAs & \0\0144 & n & $1 \times 10^{18}$\\
DBR & \01 & AlAs & \0\0538 & n & $1 \times 10^{18}$ \\
DBR & \01 & Al$_{0.5}$GaAs & \0\0144 & n & $1 \times 10^{18}$ \\
Buffer & \01 & GaAs & \01000 & n & \0$1.5 \times 10^{18}$ \\
Substrate & & GaAs & & n & \\
\br
\end{tabular}
\end{center}
\end{table}

\begin{figure}
\center \includegraphics*[width=110mm]{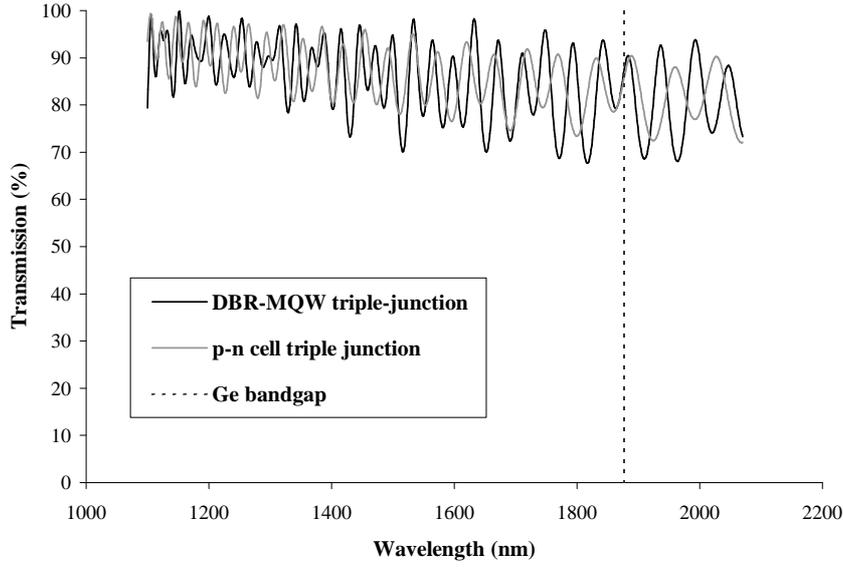}
\caption{Calculated transmission through to a Ge substrate of a
tandem cell with a GaInP top cell and as bottom cell a QWC with
DBR, compared with a tandem with a GaAs p-n bottom cell.}
\label{mqw-dbr-transmission}
\end{figure}

As the stop-band of the DBR can be made quite abrupt, most longer
wavelength light is allowed through, so that an active Ge
substrate as a third junction can be employed, which would still
produce more than 200 A/m$^2$, more than enough photocurrent
compared with the other two junctions (see below). The
transmission through to a Ge substrate of a tandem cell with a
GaInP top cell and a QWC with DBR as bottom cell has been modelled
with a multi-layer programme and is shown in
\fref{mqw-dbr-transmission} compared with a tandem having a GaAs
p-n junction as a bottom cell. As one can see the transmission of
photons of energy below the bandgap of the bottom cell is
similarly high in both cases.

In \tref{dbr-jsc} the short-circuit current density \jsc for AM0
illumination is shown for this QWC (20 QWs) with and without a
20.5 period DBR, compared with a p-n control cell
\cite{bushnell:01}. In a tandem configuration, assuming a cut-off
wavelength of 650 nm, corresponding to the bandgap of a GaInP top
cell, the p-n control cell is expected to have a typical \jsc of
158 A/m$^2$. Introducing a QWC with DBR improves \jsc by 16\% to
183 A/m$^2$, which is much better current matched to the GaInP top
cell. A commercial triple-junction cell with GaInP top cell, a
standard p-n GaAs junction and an active Ge substrate has an AM0
efficiency of 26.0\% \cite{fatemi:01}; if the GaAs junction is
replaced with a QWC with DBR the efficiency is calculated to
improve by about 3.4 percentage points or 13\% to 29.4\%.

\begin{table}
\caption{Short-circuit current densities \jsc and efficiencies for
AM0 illumination. The tandem \jsc assumes a cut-off wavelength of
650 nm. The QWC has 20 QWs and the DBR 20.5 periods. The
triple-junction efficiencies are based on a device with a GaInP
top cell and an active Ge substrate.} \label{dbr-jsc}
\begin{center}
\begin{tabular}{@{}lccc@{}}
\br  & \jsc  & Tandem \jsc  & Triple-junction   \\
Cell type & (A/m$^2$) & (A/m$^2$) & efficiency \\
\mr
QWC with DBR & 339 & 183 & 29.4\% \\
QWC without DBR & 330 & 171 & \\
p-n control &  320 & 158 & 26.0\% \cite{fatemi:01}\\
\br
\end{tabular}
\end{center}
\end{table}

\section[QWCs for TPV]{QWCs for Thermophotovoltaics}\label{sec:tpv}

Thermophotovoltaics (TPV) is the same principle as photovoltaics
but the source is at a lower temperature than the sun, typically
around 1500-2000 K instead of 6000 K, and much closer. In TPV
applications often a combustion process is used as heat source
(e.g.\ using fossil fuels or biomass), but other heat sources such
as nuclear, indirect solar or industrial high-grade waste heat can
be used too. Because the source has a lower temperature in TPV,
lower bandgap materials are required to absorb the (near-)infrared
light more efficiently. Often a selective emitter is introduced
between the source and the photovoltaic cells to obtain a narrow
band as opposed to a broad black or grey-body spectrum, increasing
the cell conversion efficiency. TPV is described in more detail
for example in references \cite{coutts:99,barnham:03}.

QWCs have advantages for TPV applications, and substantial
progress has been made in the development of QWCs for TPV focusing
on InP-based materials \cite{connolly:03}. QWCs were first
introduced for TPV applications independently by Griffin et
al~\cite{griffin:97} and Freundlich et
al~\cite{freundlich:00patent}. These InGaAs/InP QWCs indicated
better performance than InGaAs bulk cells, for example enhancing
\voc\ and improved temperature dependence. These are important
parameters in TPV, as the cells are very close to a hot source,
and the power densities are high giving rise to higher currents
which may pose series resistance problems. Further development of
QWCs has been in the quaternary system \ingaasp\ lattice-matched
to InP ($x=0.47y$), with very good material quality, and which can
be optimised for a rare-earth selective emitter Erbia having a
peak emission of about 1.5 $\mu$m, for example, but which is also
attractive for hybrid solar-TPV applications \cite{rohr:nrel:98}.

Many TPV systems operate at temperatures more suitable for longer
wavelength selective emitters such as Thulia and Holmia with peak
emissions of about 1.7 and 1.95 $\mu$m respectively. However, the
smallest bandgap achievable with lattice-matched material on InP
is that of \bulkingaas\ with an absorption edge of about 1.7
$\mu$m (see \fref{Eg-a}.) Strain-compensation techniques can be
employed to extend the absorption, which are unique to QWCs
\cite{rohr:99}.

\begin{figure}
\center \includegraphics*[width=110mm]{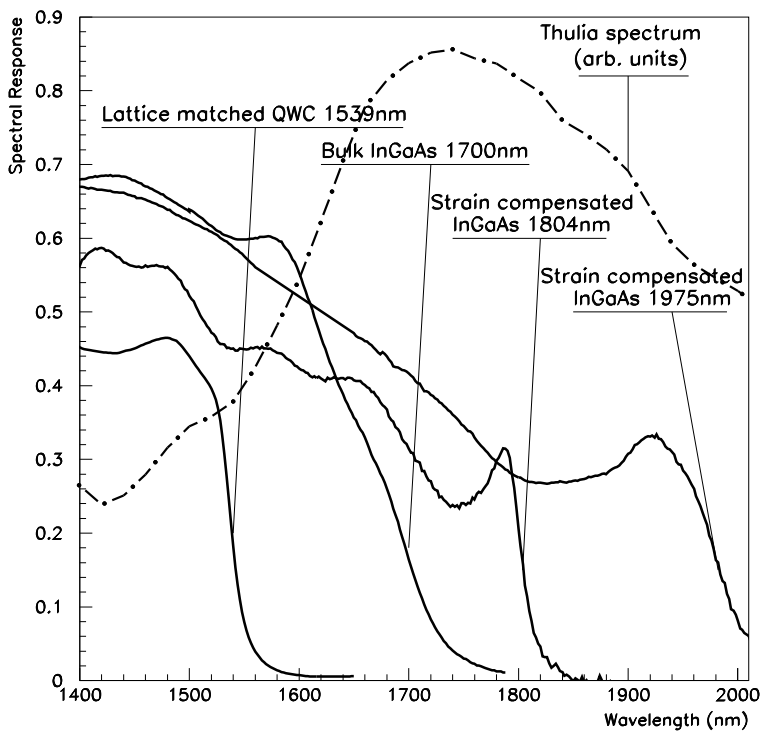}
\caption{Experimental external quantum efficiency of TPV QWC
devices (plus a bulk \bulkingaas\ device) with successively longer
absorption, including a Thulia spectrum (arbitrary units).}
\label{tpv-qe}
\end{figure}

\begin{figure}
\center \includegraphics*[width=110mm]{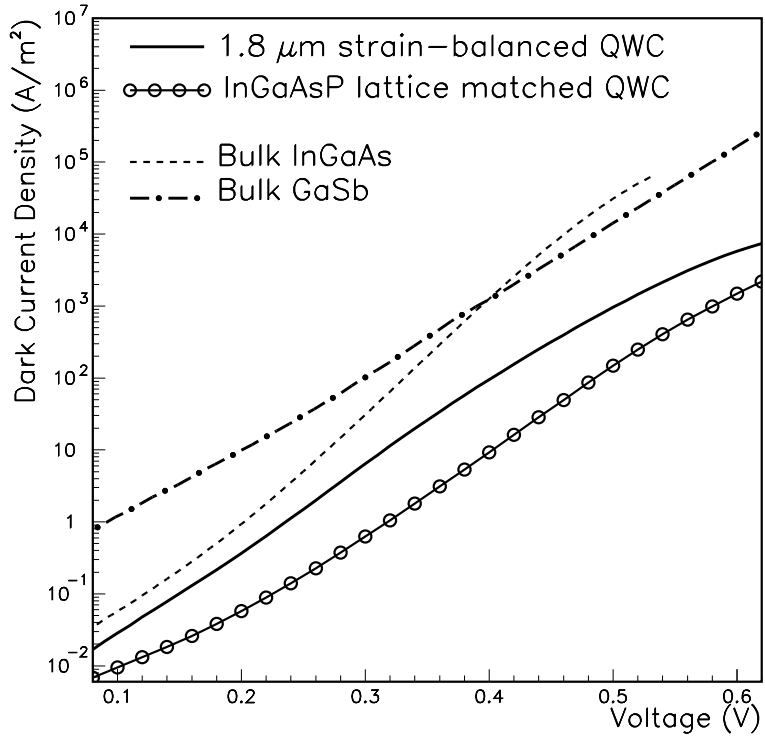}
\caption{Dark current densities of TPV cells.} \label{tpv-div}
\end{figure}

\ingaasingaas\ strain-compensated QWCs have been shown to extend
the absorption (see \fref{tpv-qe}) while retaining a dark current
similar to an \bulkingaas\ bulk cell , and lower than that of a
GaSb cell with similar bandgap (see \fref{tpv-div})
\cite{rohr:00,abbott:02}. A 40 QW \ingaasingaas\
strain-compensated QWC was designed for a Thulia emitter
\cite{abbott:tpv02}, absorbing out to 1.97 $\mu$m as shown in
\fref{tpv-qe}. Absorption beyond 2 $\mu$m has been achieved with a
2 QW cell, although in this case the collection efficiency was
incomplete except at high temperatures and reverse bias
\cite{rohr:02}.

A mirror at the back can significantly increase the contribution
of the QWs as discussed in \sref{sec:dbr}. Back reflection is
particularly important as a form of spectral control in a TPV
system, as below-bandgap photons are reflected back to the source
increasing the system efficiency. We observe a back reflectivity
of about 65\% just from a standard gold back contact
\cite{abbott:02}.

As an alternative, Si/SiGe QWCs were grown for TPV applications,
but the absorption of the SiGe QWs is very low
\cite{palfinger:02}.

\section{Conclusions}

The primary advantage of incorporating quantum wells in
photovoltaic cells is the flexibility offered by bandgap
engineering by varying QW width and composition. The use of
strain-compensation further increases this flexibility by
extending the range of materials and compositions that can be
employed to achieve absorption thresholds at lattice constants
that do not exist in bulk material. In a tandem or multi-junction
configuration QWCs allow current matching and optimising the
bandgaps for higher efficiencies.

Light trapping schemes are an important technique to boost the
quantum efficiency in the QWs. DBRs are particularly suited for
QWCs in multi-junction devices, allowing light transmission to the
lower bandgap junctions underneath.

The flexibility of materials, in particular with strain
compensation, combined with the enhanced performance of QWCs make
them especially suitable for TPV. QWCs are unique in that the
absorption can be extended to wavelengths unattainable for
lattice-matched bulk cells, while retaining a similar dark
current. Strain-compensated QWCs can be optimised for
long-wavelength selective emitters such as Thulia.

%\bibliographystyle{iopbook}
%\bibliography{master}

\end{document}